\documentclass{elsart}
\usepackage{amssymb}
\usepackage{graphics}
\usepackage{psfrag}
\usepackage{epsfig}
\journal{Physics Letters B}
\def\bea{\begin{eqnarray}}
\def\eea{\end{eqnarray}}
\def\beq{\begin{equation}}
\def\eeq{\end{equation}}
\def\nn{\nonumber}
\def\bq{\mathbf{q}}
\def\br{\mathbf{r}}
\def\bL{\mathbf{L}}
\def\bS{\mathbf{S}}
\def\bsigma{\mathbf{\sigma}}
\newcommand{\pbarp}{{\bar p p}}
\newcommand{\nbarn}{{\bar N  N}}
\newcommand{\lbarl}{{\bar \Lambda \Lambda}}
\newcommand{\lcbarlc}{\bar{\Lambda}_c^- {\Lambda}_c^+}
\begin{document}
\begin{frontmatter}
%
% PREPRINT FZJ-IKP-TH-2009-38
%
\title{The reaction $\pbarp \rightarrow \lcbarlc$ close to threshold}
\author{J. Haidenbauer$^1$ and G. Krein$^2$}

{\small $^1$Institute for Advanced Simulation and J\"ulich Center for Hadron
Physics, Forschungszentrum J\"ulich, D-52425 J\"ulich, Germany} \\
{\small $^2$Instituto de F\'{\i}sica Te\'{o}rica, Universidade Estadual
Paulista,
Rua Dr. Bento Teobaldo Ferraz, 271 - 01140-070 S\~{a}o Paulo, SP, Brazil
} \\

\begin{abstract}
Predictions for the charm-production reaction $\pbarp \to \lcbarlc$
for energies near the threshold are presented. The calculations are 
performed in a meson-exchange framework in close analogy to our earlier
study on $\pbarp \to \lbarl$ by connecting the
two processes via SU(4) symmetry. 
The obtained $\lcbarlc$ production cross sections are in the
order of 1 to 7 $\mu b$, i.e. a factor of around 10 
smaller than the corresponding cross sections for $\lbarl$.
However, they are 100 to 1000 times larger than
predictions of other model calculations in the literature.
\end{abstract}

\end{frontmatter}

\section{Introduction}
\label{sec:intro}

The study of the production of charmed hadrons in antiproton-proton ($\pbarp$)
collisions is of importance for the understanding of the strong force in the
nonperturbative regime of QCD. The FAIR project at the GSI laboratory has an 
extensive program aiming at a high-accuracy spectroscopy of charmed hadrons and at 
an investigation of their interactions with ordinary matter \cite{PANDA}. 
Presently little is known 
about such interactions and their knowledge is a prerequisite for investigating issues 
like in-medium properties of charmed hadrons, e.g. $c\bar{c}$-quarkonium dissociation 
and changes in properties of $D$ mesons due to chiral symmetry restoration effects
on the light quarks composing these mesons. Therefore there is an urgent need 
for theoretical investigations to guide such experiments.

In the present paper we concentrate on the reaction $\pbarp \rightarrow \lcbarlc$
close to its threshold. Providing predictions or simply estimations for this reaction is 
very challenging. First of all, the lack of any direct empirical information on this 
reaction makes it difficult to constrain model parameters. Second, both long-distance 
and short-distance physics are present in the reaction and this poses the question on 
the appropriate degrees of freedom to be used in order to describe it, quarks and 
gluons or mesons and baryons, or a combination of both. To the best of our knowledge, 
presently there are only four elaborate studies that consider this reaction. 
The most recent publication is 
by Goritschnig et al.~\cite{Goritschnig:2009sq}, who employ a 
quark-gluon description based on a factorization hypothesis of hard and soft 
processes. This work supersedes an earlier study by that group within
a quark-diquark picture, where already concrete predictions for the 
$\lcbarlc$ production cross section were given~\cite{Kroll:1988cd}.
In the study by Kaidalov and Volkovitsky~\cite{Kaidalov:1994mda} 
a non-perturbative quark-gluon string model is used, based on secondary Regge pole 
exchanges including absorptive corrections. On the same lines, there is the 
more recent publication by Titov and K\"ampfer~\cite{Titov:2008yf}. 

Our work here builds on the J\"ulich meson-baryon 
model~\cite{Haidenbauer:1991kt,Haidenbauer:1992wp} for the reaction 
$\pbarp \rightarrow  \lbarl$. In that model the hyperon-production reaction 
is considered within a coupled-channel approach in which 
the microscopic strangeness production process and the elastic parts of 
the interactions in the initial ($\pbarp$) and final ($\lbarl$) states are 
described by meson exchanges, while annihilation processes are accounted for 
by phenomenological optical potentials.
The elastic parts of the initial- and final-state interactions
(ISI and FSI) are $G$-parity transforms of an one-boson-exchange variant of the
Bonn $NN$ potential~\cite{obepf} and of the hyperon-nucleon model~A of 
Ref.~\cite{Holzenkamp:1989tq}, respectively. 
The model achieved a reasonably good overall description of the wealth of 
$\pbarp \to \lbarl$ data collected by the P185 collaboration at LEAR (CERN) 
on total and differential cross-sections and spin observables \cite{PS185o}
-- see also the review in Ref.~\cite{PS185}. 

The extension of the model to the charm 
sector here follows a strategy similar to our recent work on the $DN$ and ${\bar D}N$ 
interactions~\cite{Haidenbauer:2007jq,Haidenbauer:2008ff}, which used the original 
J\"ulich meson-exchange model for the $KN$ system~\cite{Buettgen:1990yw,Hoffmann:1995ie} 
and improvements from quark-gluon dynamics at short distances 
\cite{Hadjimichef:2002xe,Haidenbauer:2003rw}. 
Specifically, in the present paper we construct an extension of 
the meson-exchange model of Refs.~\cite{Haidenbauer:1991kt,Haidenbauer:1992wp}
to the reaction $\pbarp \rightarrow \lcbarlc$ assuming as a working 
hypothesis SU(4) symmetry constraints. We examine the sensitivity of the results 
to changes in the elastic and annihilation parts of the initial $\pbarp$ 
interaction and inspect the effects of the final $\lcbarlc$ interaction and of
the form factors that enter in the $\pbarp \rightarrow \lcbarlc$ transition 
potential described by $t$-channel $D$ and $D^*$ meson exchanges. In addition, 
we also investigate the effect of replacing this meson-exchange transition by a 
charm-production potential derived in a quark model. We believe this is important
for assessing uncertainties in the model, since one could easily raise the 
question on the validity of a meson-exchange description of the transition
in view of the large masses of the exchanged mesons.  

In the next Section we discuss the basic ingredients of the model and fix 
parameters by fitting differential and total inclusive $\pbarp $ cross sections. 
In Section~\ref{sec:res} we present numerical results for our predictions 
for differential and total cross sections and compare with the results available 
in the literature. A summary of our work is presented in 
Section~\ref{sec:sum}.

%%%%%%%%%%%%%%%%%%%%%%%%%%%%%%%%%%%%%%%%%%%%%%%%%%%%%%%
\section{The model}
\label{sec:model}

We will start discussing the basic ingredients of the original J\"ulich coupled channel
approach~\cite{Haidenbauer:1991kt,Haidenbauer:1992wp} that we employ here for 
the reaction $\pbarp \rightarrow \lcbarlc$. The transition amplitude is
obtained from the solution of a coupled-channel Lippmann-Schwinger equation,
\begin{eqnarray}
T(\bq',\bq,z) &=& V(\bq',\bq,z) + \int d^3q'' \, V(\bq',\bq'',z) \, G_0(\bq'',z) \, T(\bq'',\bq,z) ,
\label{T-mat}
\end{eqnarray}
where $z$ is the initial energy and $\bq'$ ($\bq$) the c.m. (center-of-mass) relative 
momentum in the initial (final) state. Here $V$ is a $2 \times 2$ matrix in channel
space containing the interaction potentials (channel $1 = p\bar{p}$, channel $2 = \lcbarlc$)
\begin{equation}
V(\bq',\bq,z) = \left(\begin{array}{cc}
  V^{11}(\bq',\bq,z) & V^{12}(\bq',\bq,z) \\
  V^{21}(\bq',\bq,z) & V^{22}(\bq',\bq,z) \\
\end{array}\right) \ , 
\label{L}
\end{equation}
and $G_0(\bq,z)$ is the propagator
\begin{equation}
G_0(\bq,z) = \left(\begin{array}{cc}
  1/(z - E^{(1)}_{\bq} + i\epsilon) & 0  \\
  0 & 1/(z - E^{(2)}_{\bq}  + i\epsilon) \\
\end{array}\right) ,
\label{G0}
\end{equation}
with $E^{(1)}_{\bq} = E_{\bq}(p) + E_{\bq}(\bar p)$ and 
$E^{(2)}_{\bq} = E_{\bq}(\Lambda_c) + E_{\bq}(\bar\Lambda_c)$. 
The diagonal potentials $V^{ii}$ are given by the 
sum of an elastic part and an annihilation part. 
 
\begin{figure}[t]
\begin{center}
\includegraphics[height=75mm,angle=-90]{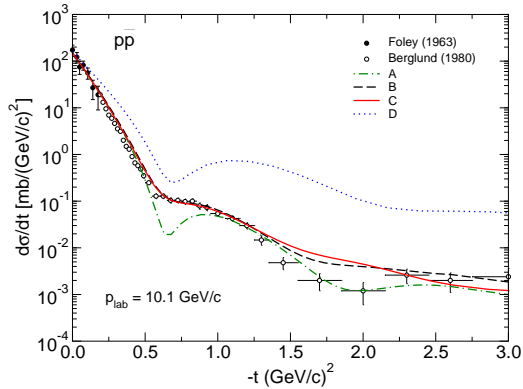}
\caption{Differential cross section for elastic $\pbarp$ scattering at
$p_{lab}$ = 10.1 GeV/c as a function of $t$. The dash-dotted curve corresponds
to a calculation where only one pion exchange is added to the optical
potential (A). The dotted curve results when the complete G-parity 
transformed OBEPF model of Ref.~\cite{obepf} is used for the elastic
part (D). The dashed and solid curves are obtained by leaving out
vector-meson exchanges (B) or by reducing the elastic part (except for
the pion exchange) to 10~\% (C), respectively. The experimental 
information is taken from Foley et al. \cite{Foley} and Berglund et 
al. \cite{Berglund}.
}
\label{pbarp}
\end{center}
\end{figure}

Though the J\"ulich group has developed rather sophisticated models of 
the $\nbarn$ interaction \cite{Hippchen,Mull} these potentials cannot be 
used anymore at
such high energies. Indeed, already in the study of $\pbarp \to \lbarl$
the elastic part of the $\nbarn$ potential was deduced (via G-parity transform)
from a simple, energy-independent one-boson-exchange $NN$ potential 
(OBEPF) and a phenomenological spin-, isospin-, and 
energy-independent optical potential of Gaussian form, 
\beq
V^{\pbarp \to \pbarp}_{opt}(r)  = (U_0 + i W_0) \, e^{- r^2/2r^2_0} ,
\label{Vaa-pp}
\eeq
was added in order to take into account annihilation. 
Now, at even much higher energies, any $NN$ potential has to be
considered as being purely phenomenological. Still we keep the longest
ranged (and model-independent) part of the elastic $\pbarp$ interaction, namely 
one-pion exchange. To it we add again an optical potential of the
form given in Eq.~(\ref{Vaa-pp}) and determine the parameters
($U_0$, $W_0$, $r_0$, cf. Table~\ref{param}) by a fit to $\nbarn$ data in the 
energy range relevant for the reaction $\pbarp \to \lcbarlc$. Fortunately, 
there are total cross sections \cite{Denisov,Foley,Astbury}
and even differential cross sections \cite{Foley,Berglund}
around $p_{lab} =$ 10 GeV/c, i.e. fairly close to the $\lcbarlc$ 
threshold which is at 10.162 GeV/c. This $\pbarp$ potential is 
called model A in the following. 
As can be seen from Table~\ref{cross} and Fig.~\ref{pbarp}
the integrated cross sections as well as the $\pbarp$ different
cross section are fairly well reproduced. 

Since one knows from studies on $\pbarp \to \lbarl$ that the magnitude
of the cross sections depends very sensitively on the ISI 
\cite{Haidenbauer:1991kt,Haidenbauer:1992wp,Kohno,Alberg}
we consider here 
several variants of the $\nbarn$ potential. First, we take the full 
G-parity transformed OBEPF interaction and add again the optical potential
(model D). 
Obviously in this case, only the total cross section can be still brought
in line with the experiment by adjusting the parameters of the optical
potential, while all other cross sections are strongly overestimated. Still
we consider this potential too with the intention that it serves as an
illustration for the uncertainties in our predictions due to the 
used $\pbarp$ interaction. Realizing that the overestimation of the 
cross sections is primarily caused by the vector-meson exchange ($\rho$,
$\omega$) contributions \cite{Machleidt} to OBEPF, we prepare two 
more potentials where we (i) leave out those vector mesons altogether 
(B) or (ii) reduce the elastic part (except for one-pion exchange) 
to 10\% (C). Within both scenarios a
rather satisfying description of the $\nbarn$ data around 10 GeV/c can be
obtained, cf. Table \ref{cross} and Fig.~\ref{pbarp}. In particular, not
only the slope but even the shoulder in the differential cross 
section is reproduced quantitatively by these two interactions. 

\begin{table}[h]
\renewcommand{\arraystretch}{1.1}
\centering
\caption{\label{param} Parameters of the phenomenological optical
potential in the $\pbarp$ channel for the four different models
described in the text. 
}
\vskip 0.2cm 
\begin{tabular}{|l|cccc|}
\hline
& A & B & C & D \\
\hline
 $U_0$ & $-48.0$ MeV & $100.5$ MeV & $72.9$ MeV & $1808$ MeV\\
$W_0$ & $-531.9$ MeV & $-529.5$ MeV & $-448.3$ MeV &$-1644$ MeV\\
$r_0$ & $0.56$ fm    & $0.56$ fm & $0.58$ fm & $0.41$ fm \\
\hline
\end{tabular}
\end{table}
\renewcommand{\arraystretch}{1.0}

\begin{table}[h]
\renewcommand{\arraystretch}{1.1}
\centering
\caption{\label{cross} Integrated $\pbarp$ cross sections at 
$p_{lab}$ = 10.1 GeV/c for the four models considered. 
The experimental value for the charge-exchange channel 
$\pbarp \to {\bar n}n $ (cex) is for $p_{lab} = 9$ GeV/c \cite{Astbury}. 
}
\vskip 0.2cm 
\begin{tabular}{|l|ccc|}
\hline
& $\sigma_{tot}$ [mb] & $\sigma_{el}$ [mb] & $\sigma_{cex}$ [mb] \\
\hline
experiment & $54.7\pm 0.60$ \cite{Denisov}  & $14.6\pm 3.3$ \cite{Foley} 
 & $0.284\pm 0.041$ \cite{Astbury} \\
\hline
A & $54.2$  & $14.4$ &  $0.20$ \\
B & $54.1$  & $14.6$ &  $0.23$ \\
C & $54.3$  & $14.2$ &  $0.42$ \\
D & $54.6$  & $20.6$ &  $2.82$ \\
\hline
\end{tabular}
\end{table}
\renewcommand{\arraystretch}{1.0}

The interaction in the final $\lcbarlc$ system is assumed to be the
same as the one in $\lbarl$. Specifically, this means that the 
elastic part of the interaction is given by the isospin-zero $\sigma$ and $\omega$
exchanges with coupling constants taken from the hyperon-nucleon 
model~A of Ref.~\cite{Holzenkamp:1989tq}, while the annihilation part is again
parameterized by an optical potential which contains, however, spin-orbit and tensor 
components in addition to a central component \cite{Haidenbauer:1991kt}:
\bea
V^{\lcbarlc\to\lcbarlc}_{opt}(r)  &=& \bigl[U_c + i W_c + (U_{LS} + i W_{LS}) \, \bL\cdot\bS 
\nn \\
&& + \, (U_t + i W_t) \, \bsigma_{\Lambda_c}\cdot\br \, \bsigma_{\bar\Lambda_c}\cdot\br\bigr]  
\, e^{- r^2/2r^2_0} .
\label{Vaa-ll}
\eea
The free parameters in the optical $\lbarl$ potential were determined in 
Ref.~\cite{Haidenbauer:1991kt} by a fit to data on total and differential cross sections, and
analyzing power for $\pbarp \to \lbarl$. Clearly, there are no reasons to believe that 
the $\lcbarlc$ interaction will be the same on a quantitative level. 
But we expect that at least the bulk properties are similar. 
Specifically, in both cases near threshold the interactions
will be govered by strong annihilation processes. We will investigate the role played
by the FSI for our results by simply switching it off. 
We use the parameters as given in Table II of Ref.~\cite{Haidenbauer:1992wp}.
 
In principle, the parameters of $V^{\pbarp\to\pbarp}_{opt}$ can only be determined together with
those for $\lcbarlc$ because we consider a coupled-channels problem. However, like in
$\lbarl$ the branching ratio $\pbarp \to \lcbarlc$ is so small that the effect of the
$\lcbarlc$ channel on the diagonal $\pbarp$ $T$-matrix can be savely neglected. 

The transition potential from $\pbarp$ to $\lcbarlc$ is given by $t$-channel $D$ and $D^*$ 
exchanges and their explicit expressions are the same as for $K$ and $K^*$ as given in
Ref.~\cite{Holzenkamp:1989tq}. 
They are of the generic form 
\beq
V^{ p\bar{p} \rightarrow \lcbarlc} (t) 
\sim \sum_{M=D,D^*} g^2_{N\Lambda_c M} 
\frac{F^2_{ N\Lambda_c M}(t)  }{t-m^2_M} ,
\label{Vtrans}
\eeq
where $g_{N\Lambda_c M}$ are coupling constants and $F^2_{ N\Lambda_c M}(t)$
are form factors. Under the assumption of $SU(4)$ symmetry the coupling constants
are the same as in the corresponding exchanges in $\pbarp \to \lbarl$. However, 
the cutoff masses in the vertex form factors cannot be taken over, because the
masses of the exchanged particles are now much larger. We use here a monopole
form factor with a cutoff mass $\Lambda$ of 3 GeV, at the $N\Lambda_cD$ as well 
as at the $N\Lambda_cD^*$ vertex but we will explore the sensitivity of the 
results to variations of the cutoff mass. 

To examine further the uncertainties, as an alternative we consider here 
also a charm-production potential inspired by quark-gluon dynamics. There is
a large literature associated with strange-hadron production in $p \bar p$ 
reactions, the best known works are those of Kohno and Weise~\cite{Kohno}, Furui and 
Faessler~\cite{Furui}, Burkardt and Dillig~\cite{Burkardt}, Roberts~\cite{Roberts} 
and Alberg, Henley and Wilets~\cite{Alberg}  -- a more complete list of references 
can be found in the review of Ref.~\cite{PS185}. In the present study we 
adopt the interaction proposed by Kohno and Weise, derived in the 
so-called $^3S_1$ mechanism of a constituent quark model. In this model the 
$s\bar s$ pair in the final state is created from an initial $u\bar u$ or $d\bar d$ 
pair via $s$-channel gluon exchange. After quark degrees-of-freedom are integrated 
out the potential has the form~\cite{Kohno}:
\beq
{V}^{\pbarp \to \lbarl} (r) = \frac{4}{3} 4\pi \frac{\alpha}{m_G^2}
\delta_{S1}\delta_{T0}\left(\frac{3}{4\pi\langle {r}^2\rangle }\right)^{3/2}
\times \exp (-3{r^2}/(4\langle {r}^2 \rangle)) \ . 
\label{QM-pot}
\eeq
Here ${\alpha}/{m_G^2}$ is an effective (quark-gluon) coupling strength,
$\langle {r}^2 \rangle$ is the mean square radius associated with the 
quark distribution in the {$p$} or $\Lambda$, and $S$ and $T$ are
the total spin and isospin in the $\pbarp$ system. 
This simple potential has actually a very modern form, namely that of
a contact term, though smeared out by the quark distribution. The
effective coupling strength is practically a free parameter that was
fixed by a fit to the $\pbarp \to \lbarl$ data \cite{Haidenbauer:1992wp}.
But it depends implicitly on the effective gluon propagator, i.e. on the 
square of the energy transfer from initial to final quark pair, 
cf. Refs.~\cite{Furui,Burkardt,Kohno1}. Heuristically this energy transfer corresponds 
roughly to the masses of the produced constituent quarks, i.e. 
$m_G \approx 2m_q$, we expect the effective coupling strength 
${\alpha}/{m_G^2}$ for charm production to be reduced by the ratio 
of the constituent quark masses of the strange and 
the charmed quark squared, $(m_s / m_c)^2 \approx$ (550 MeV / 1600 MeV)$^2$ 
$\approx$ 1/9 as compared to the one for $\pbarp \to \lbarl$. 
Note that a different suppression factor arises for the $^3P_0$ quark-antiquark 
annihilation vertex, considered by Alberg et al. in their study of 
$\pbarp \to \lbarl$ \cite{Alberg}. In this case the amplitude scales with
$1/m_q$ so that the effective strength of the transition potential is
reduced by $m_s / m_c \approx$ 1/3 only when going
from strangeness to charm production. 

%%%%%%%%%%%%%%%%%%%%%%%%%%%%%%%%%%%%%%%%%%%%%%%%%%%%%%
\section{Results}
\label{sec:res}

\begin{figure}[t]
\begin{center}
\includegraphics[height=75mm,angle=-90]{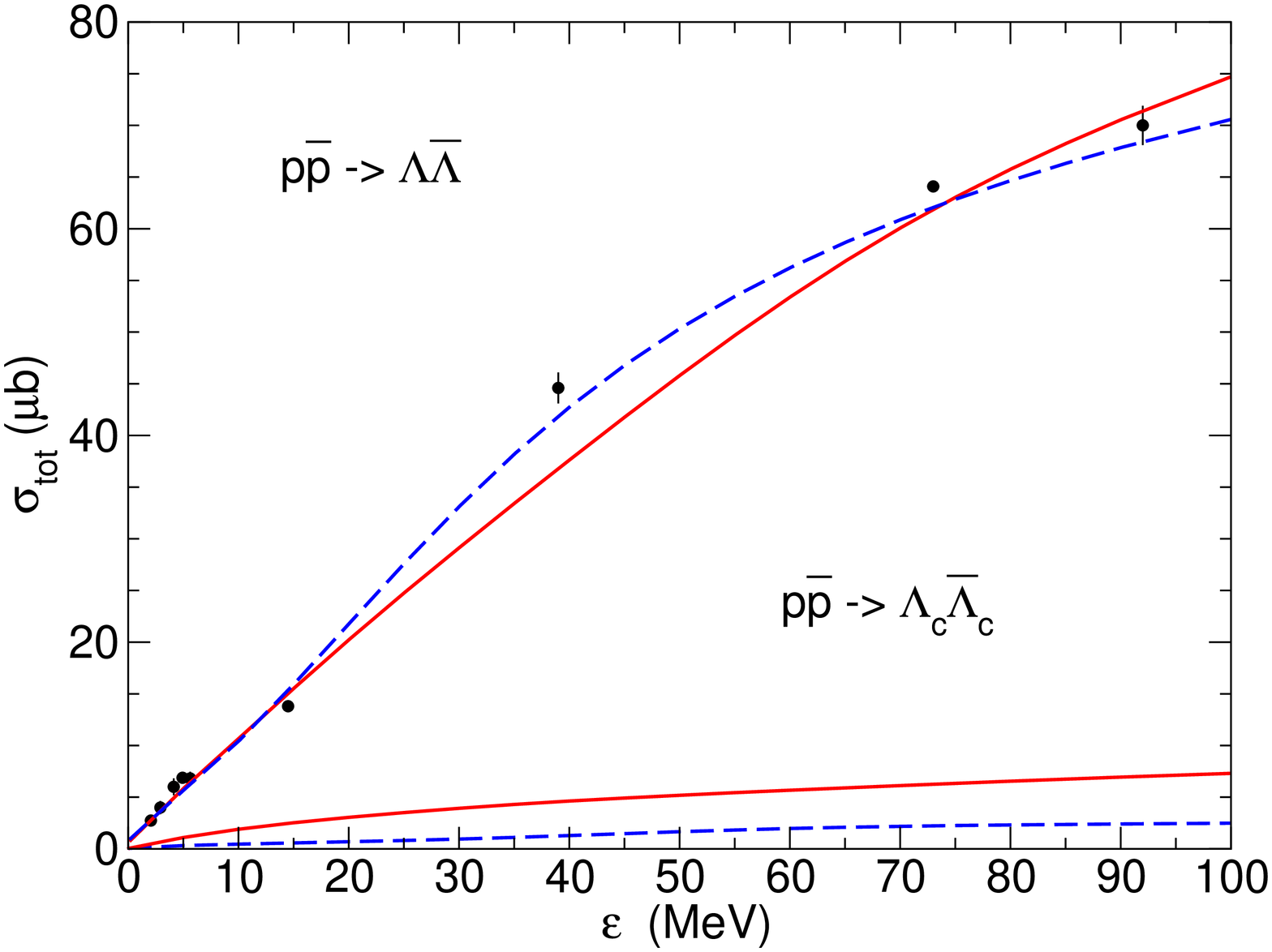}
\caption{Total reaction cross sections for 
$\pbarp \to \lbarl$ and $\pbarp \to \lcbarlc$ 
as a function of the excess energy $\epsilon$.
The results for $\pbarp \to \lbarl$ (upper curves)
are taken from our work \cite{Haidenbauer:1992wp}.
The solid curves are results for the meson-exchange
transition potential while the dashed curves 
correspond to quark-gluon dynamics. The $\pbarp \to \lcbarlc$
results are obtained with the $\pbarp$ interaction C. 
}
\label{fig:2}
\end{center}
\end{figure}

Our predictions for the total reaction cross section 
for $p\bar{p} \to \lcbarlc$ are presented
in Fig.~\ref{fig:2}. The cross section is shown as a function
of the excess energy $\epsilon = \sqrt{s} - m_{\Lambda_c}
- m_{\bar\Lambda_c}$ so that we can compare it with the one 
for $p\bar{p} \to \lbarl$ at the corresponding 
$\epsilon$. The curve in Fig.~\ref{fig:2} correspond to 
the $\pbarp$ model C, for which we obtained the largest results. 

Obviously, and as expected, the cross section for $\lcbarlc$
production is smaller than the one for $\lbarl$. But the
difference is about one order of magnitude only. Indeed, one
can get a very rough estimation for the reduction just by 
considering the following: When going from $\lbarl$ to $\lcbarlc$ 
within the meson-exchange picture the main change
occurs in the meson propagators of the transition potentials where 
the masses of the strange mesons ($K$, $K^*$) are replaced by
those of the charmed mesons ($D$, $D^*$), 
$m_{M_s} \rightarrow m_{M_c}$, cf. Eq.~(\ref{Vtrans}). 
All coupling constants remain the same because of the assumed 
SU(4) symmetry. Thus, the ratio of the transition potentials is 
then given roughly by 
$V^{\pbarp \to \lcbarlc} / V^{\pbarp \to \lbarl}
\approx m_{M_s}^2 / m_{M_c}^2 \approx {1/4}$
so that one expects the cross section to be smaller by
a factor of around 16. Our explicit calculation is pretty
much in line with this admittedly rather qualitative 
estimation. 

\begin{figure}[t]
\begin{center}
\includegraphics[height=65mm,angle=-90]{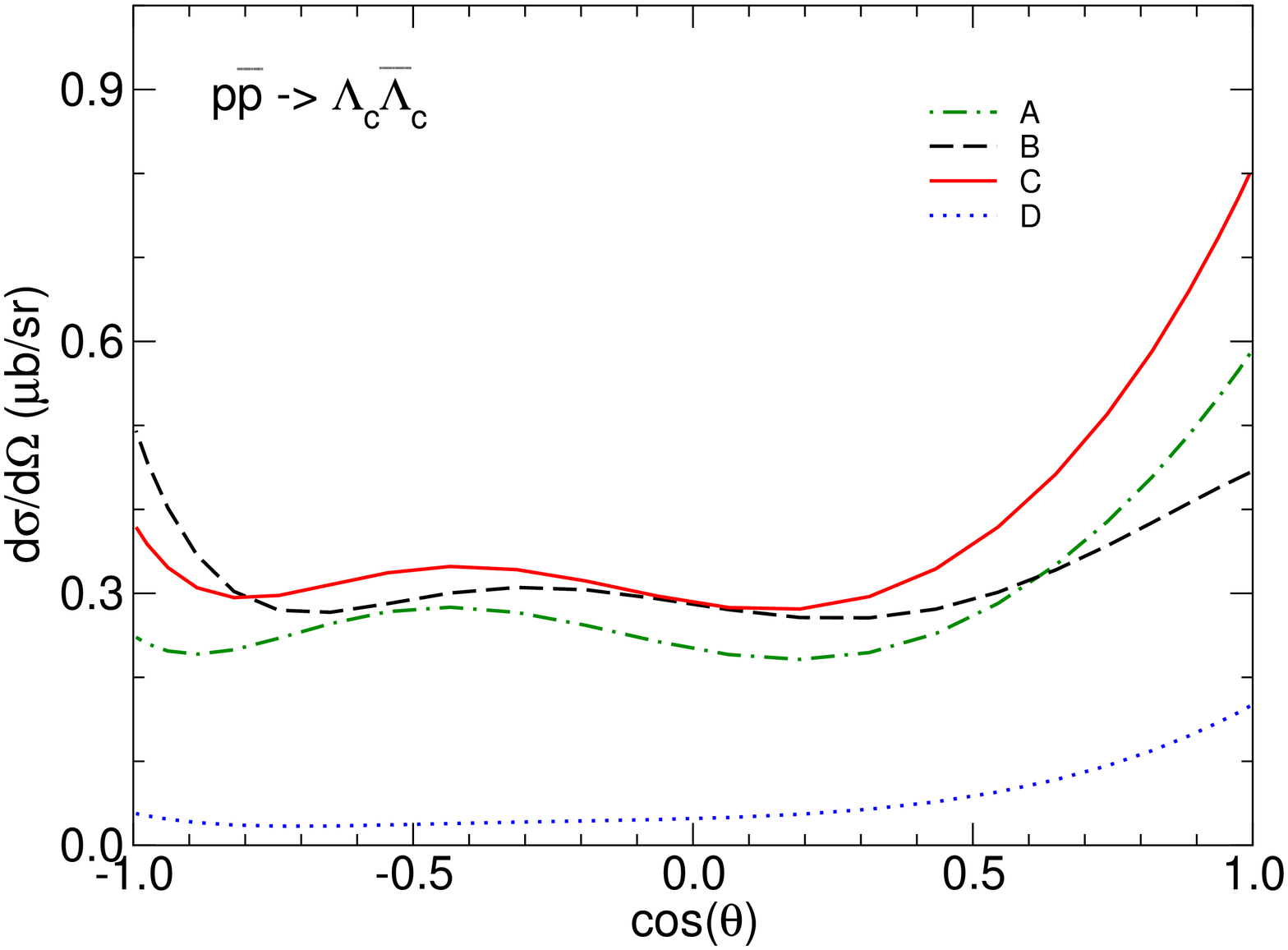}
\includegraphics[height=65mm,angle=-90]{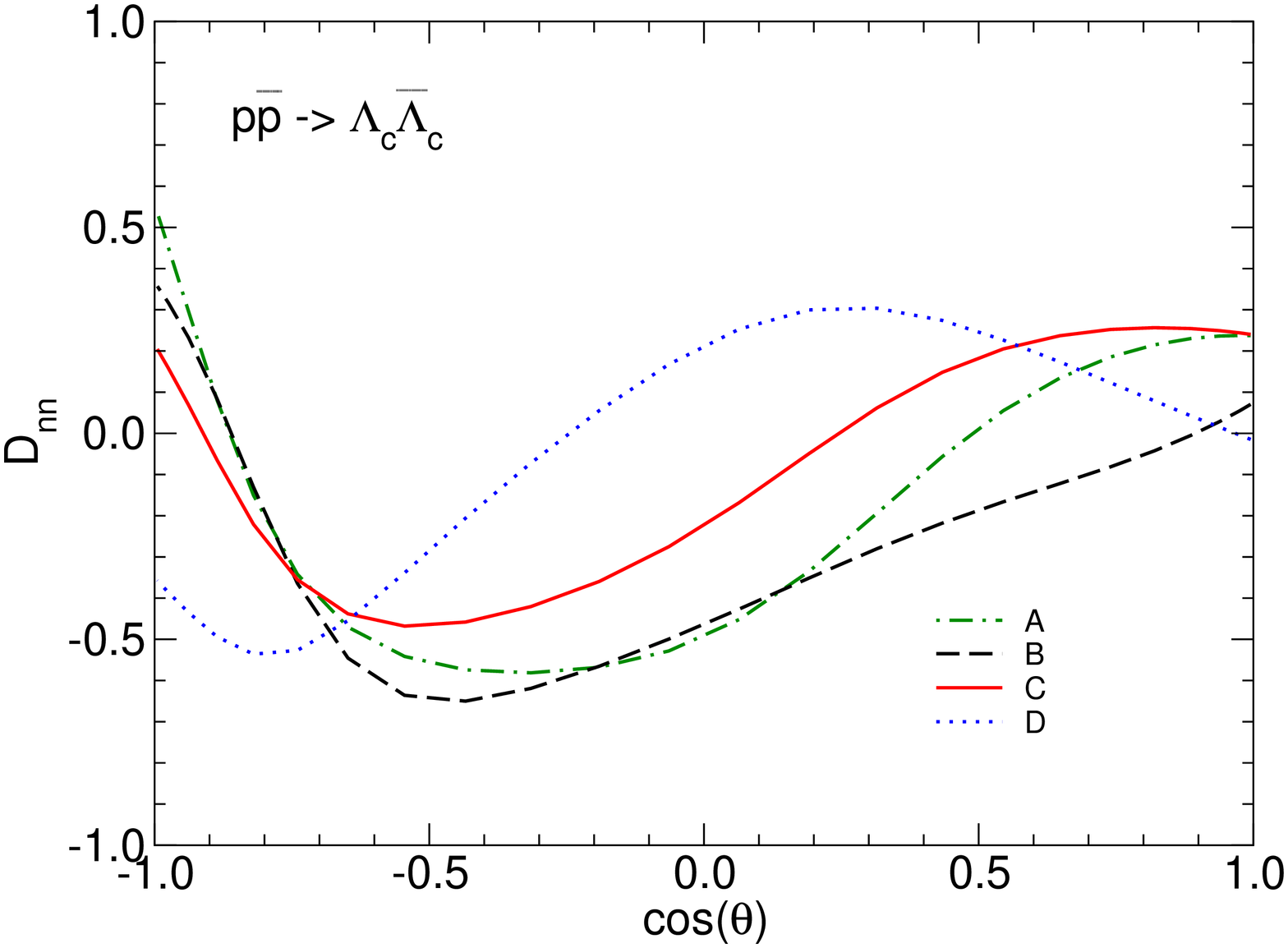}

\includegraphics[height=65mm,angle=-90]{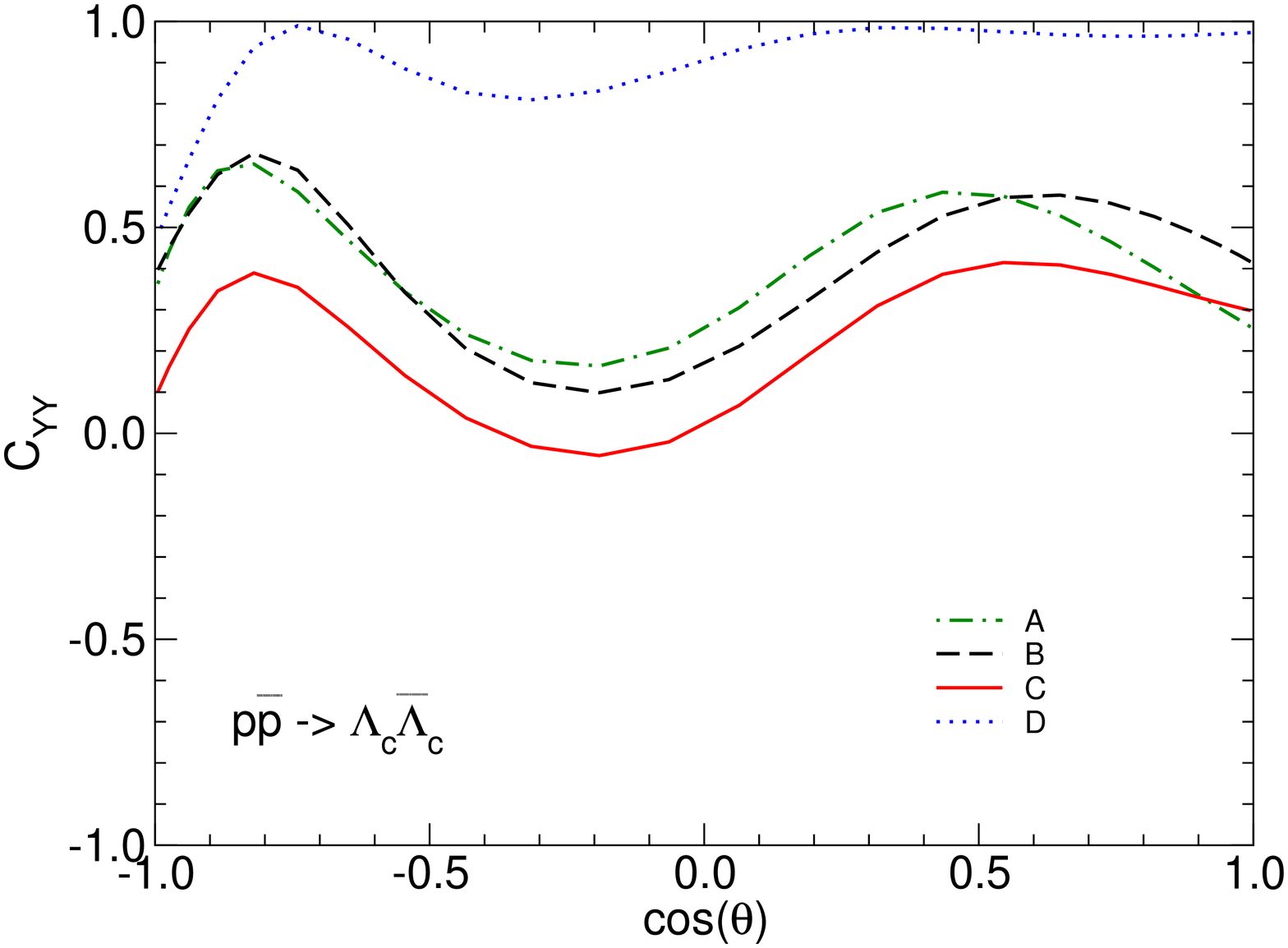}
\includegraphics[height=65mm,angle=-90]{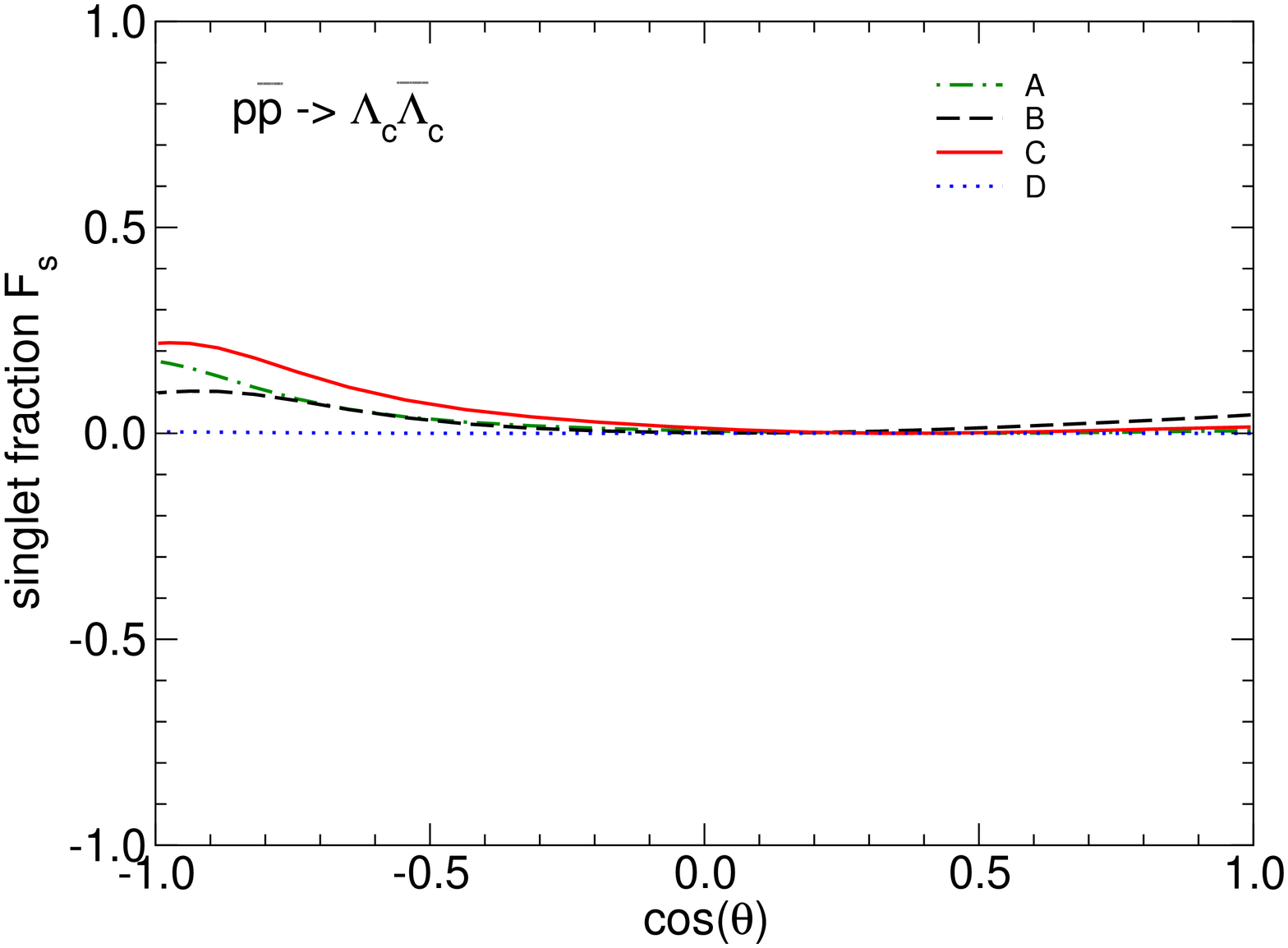}
\caption{Differential cross section, singlet fraction,
depolarization $D_{nn}$ and spin-correlation parameter
$C_{yy}$ for $\pbarp \to \lcbarlc$ 
at $p_{lab} = 10.343$ GeV/c ($\epsilon = 40$ MeV)
as a function of the the c.m. scattering angle.
The curves correspond to different ISI as indicated
in the figure. The meson-exchange potential is used
for the transition interaction. 
}
\label{fig:3}
\end{center}
\end{figure}

Let us now discuss the sensitivity of the results to the
various ingredients of our model calculation. The most
crucial component is certainly the $\pbarp$ interaction in
the initial state. Without it, and specifically in Born
approximation, we would get cross sections that are more
than a factor 100 larger. This is no surprise because the
very same gross overestimation in the Born approximation 
occurs for $\pbarp \to \lbarl$ 
\cite{Haidenbauer:1991kt,Haidenbauer:1992wp,Kohno,Alberg}
but also for other exclusive $\pbarp$ annihilation channels 
like $\pbarp \to \pi\pi$, $\pbarp \to K\bar K$, etc. 
\cite{Hippchen,Mull}, if one does not take into account the ISI. 
However, with the ISI included, it is reassuring to see that the
variations of the predicted $\pbarp \to \lcbarlc$ cross 
section for the different $\pbarp$ potentials we have prepared
is fairly small. To be concrete, the cross sections at 
$\epsilon =$ 100 MeV are 5.8, 6.3, and 7.3 $\mu b$ for the
interactions A, B, and C, respectively. 
Only the result based on the full G-parity transformed OBEPF
potential differs more
significantly. Here we get a cross section that is with 0.8 $\mu b$ 
a factor 10 smaller than the other values. 
The suppression for this $\pbarp$ model is presumably due to the fact
that it yields a $\pbarp$ elastic cross section that is much larger
than the ones of the other potentials considered, and actually in
disagreement with experimental information. Still, the variation
of one order of magnitude may be considered as a realistic estimation
of the uncertainty due to the ISI. 

Once the ISI is included our results turned out to be rather
insensitive to the final $\lcbarlc$ interaction. Even when we leave
it out altogether the cross sections do not change (decrease) by
more than 10-15 \%. 
We also considered variations of the cutoff mass at the
$N\Lambda_c D$ and $N\Lambda_c D^*$ vertices. For the results 
discussed above a value of $\Lambda =$ 3 GeV has been used. When
reducing this value to 2.5 GeV the cross sections at $\epsilon =$ 
100 MeV drop by roughly a factor 3. Of course, employing even smaller
cutoff masses would further decrease the cross sections. However, 
since the exchanged mesons have a mass of around 1.9 to 2 GeV
we consider values below 2.5 GeV as being not really realistic. 

We display here also the results based on an adaption of the $^3S_1$ 
quark-gluon transition mechanism of Ref.~\cite{Kohno}.
We scale the effective coupling strength $\alpha/m^2_G = 0.25~{\rm fm}^2$,
fixed in our study of $\pbarp \to \lbarl$~\cite{Haidenbauer:1992wp},
with $ (m_c / m_s)^2$ using the constituent quark masses 
$m_s = 550~{\rm MeV}$ and $m_c = 1600~{\rm MeV}$, i.e. the same 
values as employed in our previous works in 
Refs.~\cite{Haidenbauer:2007jq,Haidenbauer:2008ff}. 
As expected, we obtain cross sections that are of the same magnitude as 
those predicted
in the meson-exchange picture though roughly a factor three smaller, 
cf. the dashed line in Fig.~\ref{fig:2}. 
In principle the factor 
$\langle r^2 \rangle$ in Eq.~(\ref{QM-pot}) also changes when replacing 
$m_s$ by $m_c$, since it gives the size of the quark distribution of
the overlapping hadrons -- $\langle r^2 \rangle$ decreases as $m_s$ is
replaced by $m_c$. We have not changed this factor in the present 
work, since we expect this change to be much smaller than the 
change in the square of the energy transfer $m^2_G$, since hadron 
sizes do not scale as simple powers of quark masses and so the overlap 
between hadron wave functions is not expected to change drastically 
from $\Lambda$ to $\Lambda_c$ -- for explicit numbers for the size 
parameters of harmonic oscillator wave functions for $N$ and $\Lambda_c$ 
see Ref.~\cite{Hilbert:2007hc}.  
The size parameter we use is $\langle r^2 \rangle^{1/2} = 0.55~{\rm fm}$.

Predictions for differential cross sections, the singlet fraction $F_s$,
the depolarization $D_{nn}$ and the spin-correlation parameter $C_{yy}$ 
(cf. Ref.~\cite{Haidenbauer:1992wp} for definitions) at
$\epsilon =$ 40 MeV are presented in Fig.~\ref{fig:3} for all 
four ISI considered. Note that the singlet fraction is defined by 
$F_s = \frac{1}{4}(1 -\langle \vec \sigma_1\cdot \vec \sigma_2\rangle)
= \frac{1}{4}(1 + C_{xx} - C_{yy} + C_{zz})$ \cite{Haidenbauer:1991kt}.
It is zero if the $\lcbarlc$ pair is produced purely in a
triplet state. Indeed in case of $\pbarp \to \lbarl$ it was found
experimentally that the singlet fraction is close to zero \cite{PS185o}
and this feature was reproduced by meson-exchange potentials 
\cite{Haidenbauer:1991kt} as well
as by transition interactions based on quark-gluon dynamics
\cite{Haidenbauer:1992wp,Kohno,Alberg}.
In the 
former it is due to a strong tensor force generated by the combined
$K+K^*$ exchange which leads to a dominance of transitions of the
form $L_{\lbarl} = L_{\pbarp}-2$ where $L$ refers to the orbital
angular momentum. As can be seen from Fig.~\ref{fig:3}, the very
same feature is predicted also for the reaction $\pbarp \to \lcbarlc$.
The differential cross section exhibits a peak in forward direction,
similar to what has been also observed for $\pbarp \to \lbarl$
\cite{Haidenbauer:1991kt,Haidenbauer:1992wp,PS185o}.

\begin{figure}[t]
\begin{center}
\includegraphics[height=75mm,angle=-90]{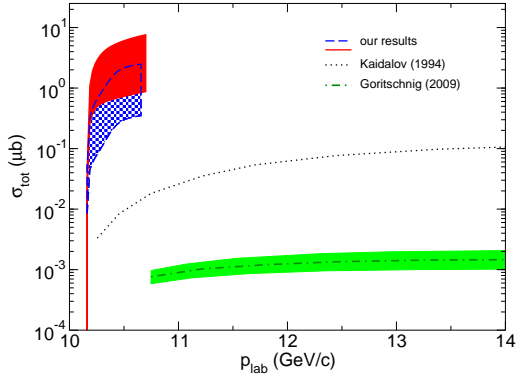}
\caption{Total reaction cross sections for 
$\pbarp  \to \lcbarlc$ as a function of the excess energy $p_{lab}$.
The dark (red) shaded band (blue grid) is the prediction of our 
meson-exchange (quark-gluon) transition potential. 
The dotted curve is the result from Ref.~\cite{Kaidalov:1994mda}
while the dash-dotted curve and the corresponding (green) band is from 
Ref.~\cite{Goritschnig:2009sq}. 
}
\label{fig:4}
\end{center}
\end{figure}

Let us now compare our predictions with those by other groups. 
This is done in Fig.~\ref{fig:4}. 
Goritschnig et al.~\cite{Goritschnig:2009sq} as well as
Kaidalov and Volkovitsky~\cite{Kaidalov:1994mda} have presented
explicit results in their publications and we reproduce them
in Fig.~\ref{fig:4} to facilitate a comparison. 
Our results are shown as dark (red) shaded band (and grid) in order 
to reflect the variation of the predictions when the four 
different ISI are used. 
It is remarkable that our results differ drastically 
from those of the preceeding works. Specifically, our
cross sections are a factor 1000 larger than those given
by Goritschnig et al. and they are still about 100 times
larger than the ones by Kaidalov and Volkovitsky. Thus,
even when considering the variation of about a factor
ten due to the ISI that we see in our results and the
uncertainties due to the unconstrained FSI and form factors
in the transition potential that amount to roughly a factor three,
we are faced with an impressive qualitative difference. 

\section{Summary}
\label{sec:sum}

In this paper we presented predictions for the charm-production
reaction $\pbarp \to \lcbarlc$. The calculations were performed
in the meson-exchange framework in close analogy to our earlier
study on $\pbarp \to \lbarl$ by connecting the
two processes via SU(4) symmetry. The interaction in
the inital $\pbarp$ interaction, which plays a crucial role for
the quantiative predictions and which is now needed at a much 
higher energy, is re-adjusted so that available $\pbarp$ scattering
data in the relevant energy range are reproduced. 

The obtained $\lcbarlc$ production cross sections are in the
order of 1 to 7 $\mu b$. Thus, they are just about a factor 10
smaller than the corresponding cross sections for $\lbarl$.
The reduction is in line with a naive estimation based on
the mass ratio of strange ($K, \ K^*$) versus charmed ($D, \ D^*$) 
mesons. The exchange of those mesons governs the range of the 
forces that are responsible for producing a $\Lambda$ or a 
$\Lambda_c$, respectively, in the meson-exchange picture. 

Interestingly, other model calculations in the literature
predict $\lcbarlc$ production cross sections that are 100
to 1000 times smaller than what we found. Since the 
approaches used in
those calculations are rather different from ours it is
practically impossible to say where this drastic difference
in the predicted cross sections comes from. Hence, a 
discrimination between those scenarios appears to be possible only
when the reaction $\pbarp \to \lcbarlc$ will be examined in a 
future experiment at FAIR. 
Of course, if our predictions turn out to be closer to
reality than the others, it will be much easier to perform 
a pertinent experiment at FAIR. Moreover, $\Lambda_c$'s and 
$\bar \Lambda_c$'s can then be produced more copiously so that 
also any other experiment that requires charmed baryons as a 
probe will be much more promising. 

The model presented here leaves room for improvements in several
directions. In particular, it would be important to extend the
simple picture of $s$-channel $q\bar q$ annihilation followed 
by $c \bar c$ creation for the transition amplitude in the quark 
model as used here. An interesting possibility is the one pursued 
in Ref.~\cite{Roberts} for $\pbarp \to \lbarl$ where interacting 
many-quark intermediate states are considered in the process.

\vskip 0.2cm 
\noindent
{\bf Acknowledgments}
We acknowledge stimulating discussions with Tord Johansson 
and Wolfgang Schweiger on the topic of this paper. We thank 
Ulf-G. Mei{\ss}ner for his careful reading of our manuscript.
Work partially supported by CNPq and FAPESP (Brazilian agencies).  

%
%%%%%%%%%%%%%%%%%%%%%%%%%%%%%%%%%%%%%%%%%%%%%%%%%%%%%%

\end{document}